\documentclass[prl,aps,twocolumn,showpacs,groupaddress,footnoteinbib]{revtex4-1}

\usepackage{graphicx}
\usepackage{dcolumn}
\usepackage{bm}
\usepackage{amssymb,amsmath}
\usepackage{sidecap}
\usepackage{wrapfig}
\usepackage[bookmarks]{hyperref}
\usepackage{natbib}

\usepackage{graphicx}
\usepackage{leftidx}
\usepackage{color}

\usepackage{bbold} 

\usepackage{wasysym} 

\newcommand{\be}{\begin{equation}}
\newcommand{\ee}{\end{equation}}
\newcommand{\ba}{\begin{align}}
\newcommand{\ea}{\end{align}}
\newcommand{\sysb}{\left\{\begin{array}}
\newcommand{\syse}{\end{array}\right.}
\newcommand{\baa}{\begin{array}}
\newcommand{\eaa}{\end{array}}

\newcommand{\matb}{\left(\begin{array}}
\newcommand{\mate}{\end{array}\right)}

\newcommand{\ran}{\right\rangle}
\newcommand{\ket}[1]{\left| #1 \ran}

\definecolor{myblue}{rgb}{0,0,0.75}

\newcommand{\Hs}{H}
\newcommand{\rs}{\rho}

\begin{document}

\title{Role of interactions in a dissipative many-body localized system}

\author{Benjamin Everest, Igor Lesanovsky, Juan P. Garrahan, Emanuele Levi}
\affiliation{School of Physics and Astronomy, University of Nottingham, Nottingham, NG7 2RD, UK}

\begin{abstract}
Recent experimental and theoretical efforts have focused on the effect of dissipation on quantum many-body systems in their many-body localized (MBL) phase.  While in the presence of dephasing noise such systems reach a unique ergodic state, their dynamics is characterized by slow relaxation manifested in non-exponential decay of self-correlations.  Here we shed light on a currently much debated issue, namely the role of interactions for this relaxation dynamics.  
We focus on the experimentally relevant situation of the evolution from an initial charge density wave in the presence of strong dephasing noise. 
We find a crossover from a regime dominated by disorder to a regime dominated by interactions, with a concomitant change of time correlators from stretched exponential to compressed exponential form. 
The strongly interacting regime can be explained in terms of nucleation and growth dynamics of relaxing regions - reminiscent of the kinetics of crystallization in soft matter systems - and should be observable experimentally.
This interaction-driven crossover suggests that the competition between interactions and noise give rise to a much richer structure of the MBL phase than anticipated so far.
\end{abstract}

\maketitle


\textit{Introduction.---} 
Many-body quantum systems in the presence of quenched disorder undergo a transition between an ergodic phase and a many-body localized (MBL) phase \cite{Altshuler1997,Basko2006,Gornyi2005,Oganesyan2007,Pal2010}.
While the transport properties of the MBL phase are still debated \cite{DeRoeck2016} it is generally accepted that it is characterized by a slow growth of entanglement entropy \cite{Znidaric2008,Bardarson2012,Serbyn2013,Luitz2015}, and ergodicity breaking which has been observed in numerical studies \cite{Pal2010,DeLuca2013,Ponte2015} and experiments \cite{Schreiber2015,Smith2015,Choi2016,Bordia2016}. The non-ergodic aspects of the MBL phase were recently proposed to be reminiscent of those of glassy systems in presence of translation invariance \cite{Horssen2015,Schiulaz2015}. 

While most literature has focused on closed quantum systems, the imperfect isolation of the cold atomic ensembles used in recent experimental observations of MBL calls for a thorough understanding of the effect of dissipation on the MBL phase. 
In Ref.\ \cite{Levi2015} a chain of interacting fermions in contact with an infinite temperature dephasing bath was studied numerically.  At conditions where the closed system would be in the MBL phase, 
a slow approach to the infinite temperature state was observed in the open system. This was characterized by a stretched exponential decay of self-correlations, and thermalization on exponentially large time-scales. 
The stretched exponential behavior was confirmed analytically in \cite{Fischer2015} where an explanation of the physics in terms of a non-interacting (Anderson) system was proposed for large disorder. The same approach was used in \cite{Medvedyeva2015} where the scaling properties of the same system were studied in the large disorder limit, finding independence of the dynamics from the interactions.

A central question is therefore whether interactions play any role in the relaxation to the ergodic state due to dephasing in an otherwise MBL system.  Here we address this question by studying the dissipative dynamics of a disordered XXZ chain in its MBL phase \cite{Levi2015}.  Our main result is that contrary to what expected from the aforementioned previous studies, depending on the interaction strength the system explores two different regimes within the MBL phase.
Namely increasing interaction strength drives the dynamics of an initial spin-density wave from a regime in which relaxation is dominated by the disorder to a regime dominated by interactions. 
The observable signature of this crossover is a change of behaviour of self-correlators, from a stretched exponential to a compressed exponential dependence with time.  This latter behaviour is due to nucleation and growth of relaxing regions.  
A crossover from stretched to compressed exponential relaxation is often a manifestation of non-equilibrium and aging behaviour in soft matter and glassy systems, see for example \cite{Cipelletti2000,Cipelletti2003,Falus2006,Ruta2012}.

\textit{Model.---}
We consider a paradigmatic MBL system, namely the disordered XXZ chain which can be mapped to a system of interacting fermions with Hamiltonian
\begin{equation}
 \Hs=J\sum_{l=k}^{N}\left(c^\dagger_k c_{k+1} + c^\dagger_{k+1} c_{k}\right) + V\sum_{k=1}^N n_k n_{k+1}+\sum_{k=1}^Nh_k n_k,
 \label{eq:defHamiltonian}
\end{equation}
where we denote with $c^\dagger_k$ the fermion creation operator, with $n_k=c^\dagger_kc_k$ the number operator, and the random field $h_k \in \left[-h,h\right]$ is independently drawn for each site from a uniform distribution. This model exhibits a many-body localization transition even at infinite temperature at $h_c/J\simeq 7.2$~\cite{Pal2010,Luitz2015,Serbyn2015b}.

Following \cite{Levi2015} we couple the system to an infinite temperature Markovian dephasing bath. Considering weak coupling between the system and the bath the dynamics of the system can be described by a quantum Master equation of Lindblad form \cite{Lindblad1976,Gardiner2004}
\begin{equation}
 \dot{\rs}(t)=-i\left[\Hs,\rs(t)\right]+\gamma\sum_{k=1}^N\left[n_{k}\rs(t)n_k-\frac{1}{2}\{n_{k},\rs(t)\}\right],
 \label{eq:defMasterEquation}
\end{equation}
where $\rs$ is the system's density matrix and  $\gamma\geq0$ sets the coupling to the bath. 
The Master equation (\ref{eq:defMasterEquation}) is the simplest way of describing a system coupled to a bath, and has the advantage of being experimentally relevant, as it can be derived from microscopic principles for experiments on both cold fermionic \cite{Sarkar2014} and bosonic \cite{Pichler2010} gases in the lowest band of an optical lattice. 
The decoherence is caused by off-resonant scattering of photons forming the lattice potential, and the  dissipation rate $\gamma$ is controlled  by  the  detuning  and  the intensity  of  the trapping laser. The dynamics described by Eq. (\ref{eq:defMasterEquation}) conserves the number of fermions, and in what follows we will focus on the half-filling sector.


\textit{Rate equation description.---}
In the case of large dephasing $\gamma\gg J$ the dynamics for times $t\gg 1/\gamma$ can be efficiently described by a rate equation which describes the evolution of the diagonal elements of the density matrix $p_\alpha$ \cite{Lesanovsky2013,Cai2013,Marcuzzi2014,Everest2016}, such that the state of the system can be expressed as $\ket{p}=\sum_\alpha p_\alpha \ket{\alpha}$, where $\ket{\alpha}$ are the  $N!/(N/2)!^2$ Fock states in the half-filling sector (see also \cite{Medvedyeva2015}) and $p_\alpha$ are the associated probabilities. Here Eq. (\ref{eq:defMasterEquation}) reduces to 
\begin{equation}
\partial_\tau| p \rangle =
\sum_{k=1}^{N} \Gamma_{k} 
\left[ c_{k}^\dagger c_{k+1}+c_{k+1}^\dagger c_{k} -\mathcal{P}_k \right] | p \rangle
 \label{eq:defClassical}
\end{equation}
where $\mathcal{P}_k=n_k+n_{k+1}-2n_k n_{k+1}$. 
Eq.~(\ref{eq:defClassical}) describes classical hopping of particles on the lattice, with a rescaled time $\tau=J^2\gamma t/h^2$, cf. \cite{Fischer2015,Medvedyeva2015}.  The rate for hopping between site $k$ and $k+1$ is given by 
\begin{equation}
\Gamma_{k} = \frac{h^2}{\gamma^2+\left[V\left(n_{k+2}-n_{k-1}\right)+\Delta h_k\right]^2},
 \label{eq:ratess}
\end{equation}
where $ \Delta h_k=h_{k+1}-h_k$.
In the following we will set the energy scale by taking $\gamma=1$. The rates $\Gamma_k$ are clearly configuration-dependent, and they are illustrated in Fig.~1(a).
In the dynamics described by Eq.~(\ref{eq:defClassical}), the rates $\Gamma_k$ act as {\em kinetic constraint} \cite{Lesanovsky2013}, as often encountered in systems with complex relaxation dynamics such as glasses \cite{Garrahan2002}: the form of the rates $\Gamma_k$ does not determine the properties of the stationary state,  
but the relaxation pathways.  

Due to the random field $h_k$ the rates $\Gamma_k$ are random as well. Their probability distribution $P(\Gamma)$ depends on the strength of the interactions, and on the specific configuration under consideration.
The analytical expression of $P(\Gamma)$ is reported in the Supplementary Material \cite{SM} and it is plotted in Fig.~\ref{fig:1}(b) for various values of $V$ for an interacting configuration (namely in the cases displayed in the left column of Fig.~\ref{fig:1}(a)).
When $V<2h$, the distribution is bimodal, with a peak at $\Gamma/h^2=1$ from values of the field such that $\Delta h_k=\pm V$, [cf.\ Fig.~1(a)], and another peak at values $\Gamma/h^2\sim 4/(3h)^2$ (for $V \ll 2h$). 
The form of $P(\Gamma)$ changes qualitatively when $V>2h$: now the case $\Delta h_k=\pm V$ is not accessible, and the distribution $P(\Gamma)$ becomes unimodal retaining only the slower peak, which for $V\gg 2h$ is centered at $\Gamma\sim V^{-2}$.
The qualitative change of $P(\Gamma)$ as a function of the interaction strength already hints at the existence of different dynamical regimes within the MBL phase, which will be explored in the following.

\begin{figure}[ht]
\centering
\includegraphics[trim = 0mm 0mm 0mm 0mm, clip, width=0.75\columnwidth]{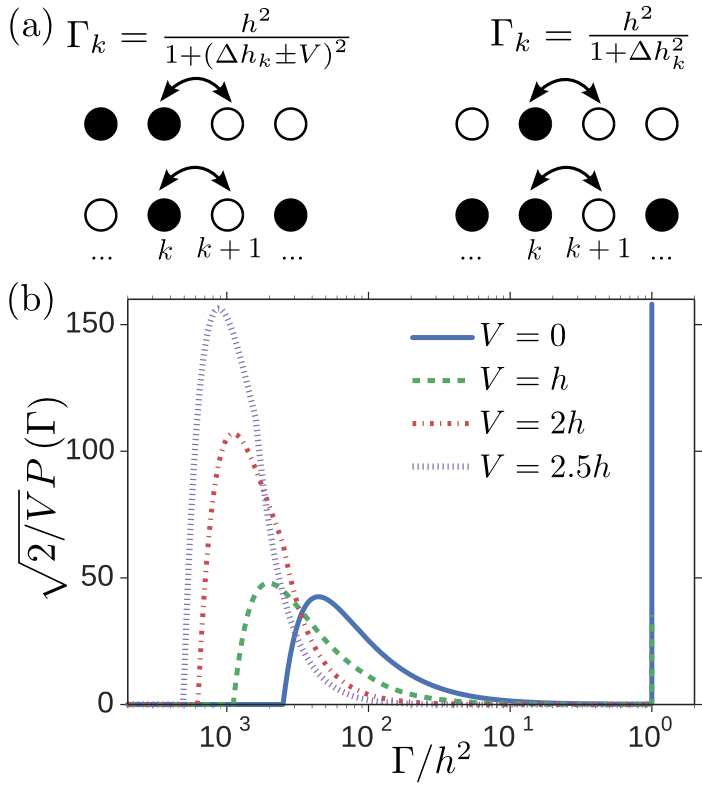}
\caption{
Panel (a) shows the dependence of the classical hopping rates $\Gamma_k$ on the configuration. 
We denote with $\bullet$ and $\circ$ respectively the occupied and empty states.
In panel (b) the normalized probability density function of rates $\sqrt{2/V}P(\Gamma)$ is displayed for different values of the interactions and $h=10$. The unnormalized distribution is depicted in the case $V=0$. 
}
\label{fig:1}
\end{figure}

\textit{Distinct dynamical regimes within the MBL phase.---}
For exploring the relaxation dynamics we focus on the case in which the initial state is the \emph{charge density wave} (CDW) state, where the corresponding probability is $|p(t=0)\rangle=\ket{\circ \bullet \circ \bullet...\circ \bullet}$, where we are denoting with $\circ,\bullet$ an empty and occupied site, respectively. 
This is in the relevant situation for recent experiments \cite{Schreiber2015,Smith2015,Choi2016,Bordia2016}, where the ergodicity properties of the system were studied via the evolution of the initial CDW quantified by the \emph{imbalance} $\mathcal{I}=(2/N)\sum_k(-1)^k n_k$. This quantity gives a direct readout of the self-correlations
\begin{equation}
 \mathcal{I}(\tau)=\frac{4}{N}\sum_{k=1}^N\left\langle n_k(\tau)n_k(0) \right\rangle-1,
 \label{eq:imbalance}
\end{equation}
and as such accounts for the ergodicity properties of the system. 

\begin{figure*}[ht]
\centering
\includegraphics[trim = 0mm 0mm 0mm 0mm, clip, width=1.91\columnwidth]{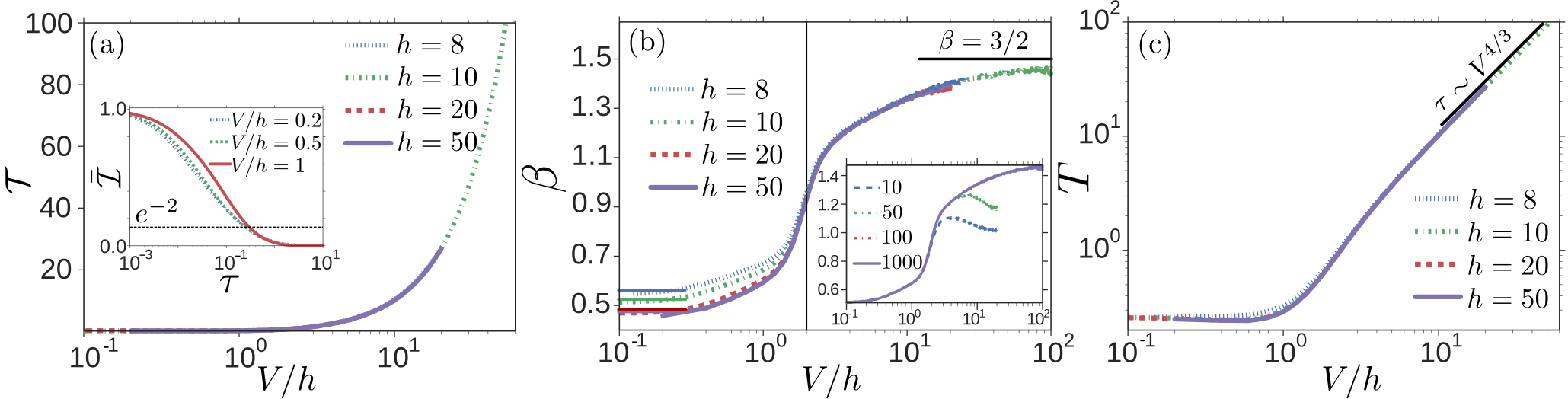}
\caption{
The data presented are for a chain of length $N=1000$ unless otherwise specified and are averaged over 10000 realizations of the disorder. Standard errors are always below the lines width.  
Panel (a) shows the dependence of the relaxation time $\mathcal{I}\left(\mathcal{T}\right)=e^{-2}$ on the interactions $V$ for various values of the disorder $h$. In the inset the relaxation dynamics is displayed for $h=10$, and different values of $V<2h$. 
In panel (b) the dependence of the exponent $\beta$ on $V$ is displayed for various values of $h$. The crossover between $\beta<1$ and $\beta>1$ at $V\simeq2h$ is highlighted by a vertical line. The analytical values obtained for large and vanishing $V$ are displayed as solid lines in the relevant regimes. In the inset the dependence of the exponent $\beta$ on $V/h$ for $h=10$ is displayed for various lengths of the chain.
Panel (c) shows the time-scale $T$ in function of $V$ for different values of $h$. The analytical value in the limit of strong interaction $T=0.32V^{4/3}$ is shown.
}
\label{fig:2}
\end{figure*} 

In Fig. \ref{fig:2} we report our results on the imbalance averaged over disorder realization $\bar{\mathcal{I}}$. The decay of $\bar{\mathcal{I}}$ becomes slower for increasing interactions. We quantify this slowing down by defining the saturation time $\mathcal{T}$ such that $\bar{\mathcal{I}}(\mathcal{T})=e^{-2}$. 
As shown in Fig. \ref{fig:2}(a) we observe two different regimes: For $V<2h$ the saturation time shows little dependence on $V$, while in contrast for $V>2h$ it increases with increasing interaction, signalling a slowdown of the dynamics. 
The inset shows that in the region $V<2h$, while $\mathcal{T}$ is approximately independent of $V$, the shape of the relaxation function depends on the on the strength of the interaction.

Our data are well fitted by the function $\bar{\mathcal{I}}(\tau)\sim \exp\left[-\left(\tau/T\right)^\beta\right]$.
This functional form is motivated by the analytical studies presented below.  
The results on the exponent $\beta$ and the time-scale $T$ are reported in Fig. \ref{fig:2}(b)-(c).
We find that at $V\simeq2h$ the relaxation of the imbalance switches from a \emph{stretched exponential} behavior ($\beta<1$) to a \emph{compressed exponential} behavior ($\beta>1$) (see Fig.~\ref{fig:2}(b)).
A finite size study for the exponent $\beta$ is shown in the inset of Fig. \ref{fig:2}(b). Although in the stretched exponential regime ($V<2h$) finite size effects have a marginal impact, in the compressed exponential regime ($V>2h$) they cause a saturation of the exponent to lower values. The origin of this behavior will become clear below.

\begin{figure}[h]
\centering
\includegraphics[trim = 0mm 0mm 0mm 0mm, clip, width=0.8\columnwidth]{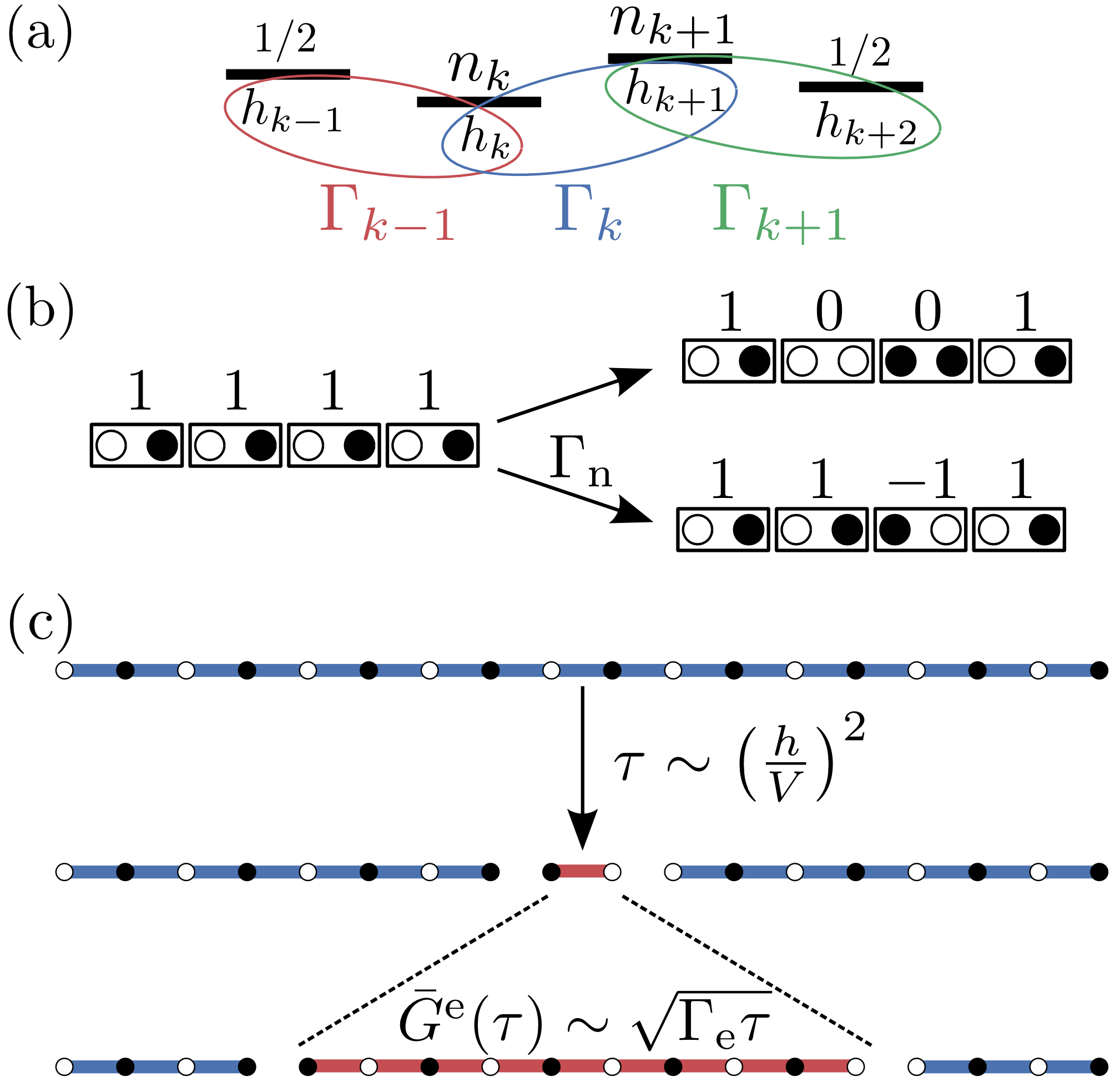}
\caption{(Colors online) 
In panel (a) the non-interacting limit is displayed: The focus is on the sites $k$ and $k+1$, while sites $k-1$ and $k+2$ serve as a bath in the relaxed state. Two contiguous rates are statistically correlated since they share the value of an on-site random field. 
In panel (b) the two possible effects of a nucleation event are displayed.
Panel (c) shows a cartoon of the nucleation and expansion of a transformed region in the strongly interacting limit. 
The transformed and untransformed regions are depicted respectively in red and blue.
}
\label{fig:3}
\end{figure}

Here we follow with explaining analytically the stretched and compressed exponential behaviors respectively for $V\ll 2h$, and $V\gg 2h$.
When $V\ll2h$ the dynamics is dominated by disorder, and we can set $V=0$. In this case the exponent $\beta$ can be understood as follows: The long time dynamics is characterized by large portions of the chain in which the system has relaxed (giving null contributions to the imbalance), with isolated non-relaxed pairs of neighboring sites corresponding to the largest $\Delta h_k$. 
The approach of $\mathcal{I}(\tau)$ to equilibrium is then determined by those sites. Their dynamics can be studied by focusing on, say, sites $k$ and $k+1$ with relaxed neighbors serving as a bath. That is, we set $n_{k'}=1/2$ for $k' > k+1$ or $k'<k$.  This setting is sketched in Fig. \ref{fig:3}(a).
The resulting effective equations for the density of the two sites under consideration are 
\begin{equation*}
 \begin{split}
  \dot{n}_k&=\Gamma_{k-1}\left(\frac{1}{2}-n_k\right)+\Gamma_{k}\left(n_{k+1}-n_k\right),\\
  \dot{n}_{k+1}&=\Gamma_{k}\left(n_{k}-n_{k+1}\right)+\Gamma_{k+1}\left(\frac{1}{2}-n_{k+1}\right).
 \end{split}
\end{equation*}
We are interested in the local imbalance $\mathcal{I}_k=n_{k+1}-n_k$ which can be obtained by integrating
\begin{equation}
\dot{\mathcal{I}}_k= -2\Gamma_{k} \mathcal{I}_k+\frac{\Gamma_{k+1}-\Gamma_{k-1}}{2}-\left(\Gamma_{k+1}n_{k+1}-\Gamma_{k-1}n_{k}\right).
 \label{eq:nonint}
\end{equation}
In this case, the rates in Eq.~(\ref{eq:ratess}) depend only on the difference of the random fields on the sites they are connecting.  The rates associated to two contiguous links (e.g., $\Gamma_{k}$ and $\Gamma_{k+1}$) are therefore not statistically independent, since they both depend on the field on the site they share, but those of links further apart are.  
When solving Eq.~(\ref{eq:nonint}) we can treat $\Gamma_{k-1}$ and $\Gamma_{k+1}$ as independent.
As an approximation we set them equal when averaging over the disorder $\Gamma_{k-1} = \Gamma_{k+1} = \Gamma'$, leading to
\begin{equation}
 \bar{\mathcal{I}}_k(\tau)=\int d\Gamma_kd\Gamma' P\left(\Gamma_k,\Gamma' \right)e^{-\left(2\Gamma_k+\Gamma'\right)\tau},
 \label{eq:joint}
\end{equation}
where $P\left(\Gamma_k,\Gamma' \right)$ is the joint probability reported in \cite{SM}.
We integrated numerically Eq.~(\ref{eq:joint}), and found a stretched exponential behavior.
In Fig.~\ref{fig:2}(b) the results on $\beta$ obtained by fitting are compared with the numerical data in the weak interaction regime, showing good agreement \footnote{The local imbalance is a good approximation of the imbalance in the regime we are considering. This is because non-relaxed links are far-apart enough to be considered independent, and they give the same average contribution.}.

In the opposite limit $V\gg2h$ on the other hand the first step ``costs'' $V+\Delta h_{k}$ when starting from the CDW state.
This sets the time-scale $\tau\sim V^2$ to observe transitions of the kind $...\circ \bullet \circ \bullet\circ\bullet...\rightarrow ...\circ \bullet \circ\circ\bullet\bullet...$.    
We will refer to these events as {\em nucleation} in the following, happening at homogeneous rate $\Gamma_{\mathrm{n}}\simeq2(h/V)^{2}$.
After a nucleation event has occurred relaxation can be achieved through transitions like $ ...\circ \bullet \circ\circ\bullet\bullet... \rightarrow ...\circ \circ \bullet\circ\bullet\bullet...$, whose rates are independent form $V$ [see e.g. Fig.~\ref{fig:1}(a)]. 
This dynamics is conveniently described by the following coarse-grained approximation.
Since the imbalance is a quantity with a period of two sites it comes natural to divide the chain in the following way: $(1,2)|(3,4)|...|(N-1,N)|$. 
We focus then on a set of new degrees of freedom labeled by the contribution that the pairs of original sites bring to the imbalance: $\ket{\circ\bullet}\rightarrow \ket{1}$, $\ket{\circ\circ}\rightarrow \ket{0}$,$\ket{\bullet\bullet}\rightarrow \ket{0}$, and $\ket{\bullet\circ}\rightarrow\ket{-1}$.
The CDW configuration corresponds to the $\ket{1,1,1,1,...,1,1}$ state, and a nucleation event creates either two 0 sites or a -1 site as shown in Fig~\ref{fig:3}(b). In what follows we will focus on the case in which two 0 sites are created. 
Once a nucleation has happened the events $\ket{0,1}\leftrightarrow \ket{-1,0}$, and $\ket{1,0}\leftrightarrow \ket{0,-1}$ are possible with the non-interacting rate $\Gamma_k=h^2/(1+\Delta h_{k}^2)$.
This pair of reversible processes implies that the $0$ sites can be treated as random walkers, which moving away from each other create a growing region of $-1$ sites.
The site dependent differences between these rates are small for the time-scales we are considering ($V \gg 2h$), and we will assume a constant rate $\Gamma_\mathrm{e}$.  
When averaged over random realizations the extension of the region between the $0$ sites expands following the law $\bar{G}_\mathrm{e}(\tau)\sim\sqrt{\Gamma_\text{e}\tau}$, contributing with net zero imbalance since the sites falling in this region are now equally likely to be a $1$ or $-1$.
This growth dynamics together with the initial nucleation events - reminiscent for example of a crystallization process - is well described by the so-called Avrami law \cite{Kolmogorov1937,Johnson1939,Avrami1939,Avrami1940,Avrami1941}. Here we give a sketch of the derivation in our case.
The average number of nucleation events up to a given time $\bar{\nu}(\tau)$ can be found by integrating $\dot{\bar{\nu}}=N \Gamma_\mathrm{n}/2$.
Not accounting for overlap of the expanded regions, the total number of transformed sites can be expressed as
\begin{equation}
\bar{\mathcal{N}}(\tau)=\int_{0}^\tau dt\quad \dot{\bar{\nu}}(t)\bar{G}_\mathrm{e}(\tau-t).
 \label{eq:totNtrans}
\end{equation}
This dynamics is sketched in Fig.~\ref{fig:3}(c) for a single expanding region of transformed sites.
Overlaps can be excluded by assuming that the increment in transformed sites $d\bar{N}_{\mathrm{tr}}$ is proportional to $d\bar{\mathcal{N}}$ multiplied by the probability of not having an already transformed site $(1-2\bar{N}_{\mathrm{tr}}/N)$, giving

\begin{equation}
 \frac{2\bar{N}_{\mathrm{tr}}(\tau)}{N}=1-\exp{\left[
 -\frac{2}{3}\sqrt{\Gamma_\mathrm{e}}\left(\frac{h}{V}\right)^2\tau^{\frac{3}{2}}\right]}.\label{eq:avrami}
\end{equation}
Initializing our dynamics in the untransformed state, with imbalance $\mathcal{I}(0)=1$, the imbalance at a given time is given by $\bar{\mathcal{I}}(\tau)=1-2\bar{N}_{\mathrm{tr}}(\tau)/N$, leading to the compressed exponential behavior with exponent $\beta=3/2$ observed in Fig.~\ref{fig:2}(b).
Equation~(\ref{eq:avrami}) also yields the functional dependence of the time-scale $T\sim (V/h)^{4/3}$ for large $V/h$, which is confirmed by our numerical study as shown in Fig.~\ref{fig:2}(c).

This picture breaks when the distance between nucleation events becomes comparable to the system length. In this case we can consider the expansion of a nucleated region as instantaneous and the imbalance as fully relaxed after a single nucleation event.
In a single realization we can then model the imbalance as $\mathcal{I}(\tau | \tau')=1-\theta(\tau-\tau')$, where $\tau'$ is the time at which the first nucleation event happens. The probability of nucleation at this time is given as $\pi(\tau')=N \exp \left( -N \tau' /V^2\right)/V^2$, such that the imbalance averaged over realizations is
\begin{equation}
\bar{\mathcal{I}}(\tau)=\int d\tau'~ \pi(\tau') \, \mathcal{I}(\tau | \tau')=\exp{\left( 
-\frac{N\tau}{V^2}
\right)
}.
 \label{eq:finitesize}
\end{equation}
This causes the saturation of the exponent $\beta$ to $1$ in the inset of Fig.~\ref{fig:2}(b) for e.g. a system of length $N=10$.

\textit{Conclusions.---} Here we have studied the effect of interactions on the dynamics of correlations of a MBL system subject to strong dephasing noise.
We found two distinct regimes, one in which the relaxation time is dominated by disorder, and one dominated by interactions.  The physical manifestation is a crossover in the relaxation of time correlators, from stretched to compressed exponential in time.  While the stretched exponential regime is expected in the dissipative MBL case with weak interactions, the existence of a compressed exponential regime is novel to the field, and we predict it should be possible to observe it in experiments at the right conditions.  
We explain the compressed exponential behaviour as a consequence of nucleation and growth dynamics of relaxed spatial regions.
The collective aspect of the nucleation and growth dynamics also implies strong finite size effects in this regime.  
This makes our rate-equation approach crucial, since it allows us to access much larger system sizes.
These phenomena should be readily observable in current experiments on MBL, which focus on large systems, and are subject to strong dephasing.

\acknowledgements
\textit{Acknowledgements.---} E.L. would like to thank M. Marcuzzi, R. Guti\'errez and J. Min\'{a}\v{r} for insightful discussions. The research leading to these results has received funding from the European Research Council under the European Union's Seventh Framework Programme (FP/2007-2013) / ERC Grant Agreement No. 335266 (ESCQUMA), and EPSRC Grant No. EP/M014266/1.









\onecolumngrid
\newpage
\appendix
\section{Supplementary Material}
\begin{figure}[ht]
\centering
\includegraphics[trim = 0mm 0mm 0mm 0mm, clip, width=0.5\textwidth]{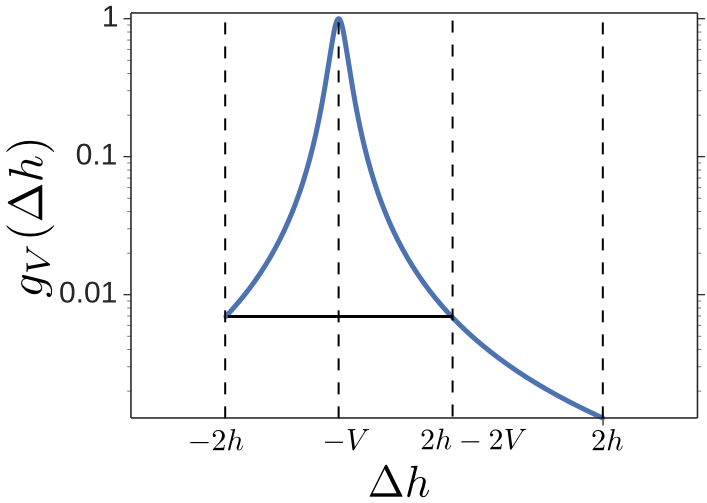}
\caption{The function $g_V(\Delta h)$ as defined in Eq.~(\ref{eq:rreate}) is shown for $h=10$, and $V=8$.}
\label{fig:1}
\end{figure}
Here we obtain analytically the distributions of rates $P(\Gamma)$ depicted in Fig.~2(b) in the main text.
Here we consider the rate associated with a fermion hopping in an ``interacting'' configuration, namely the ones depicted in the left column of Fig.~1(a) in the main text (In particular we will focus on the bottom case in the left column of Fig.~1(a), where the interaction $V$ comes with a plus sign in the rate, the other case being trivially deducible by this case). The ``non-interacting'' cases can be extracted easily from the results below by setting the interactions $V=0$.
The quantity of interest is the rate for a hopping event involving sites $k$ and $k+1$ (which we normalize by $h^2$ for simplicity), corresponding to 
\begin{equation}
\begin{split}
g_{V}(\Delta h_k)\equiv \frac{\Gamma_k}{h^2}&=\frac{1}{1+\left(\Delta h_k+ V\right)^2},\\
g_{V}^{-1}(\Gamma_k)_{\pm}&=V\pm \sqrt{\frac{1}{\Gamma_k}-1},
\end{split}
 \label{eq:rreate}
\end{equation}
where $\Delta h_k=h_{k+1}-h_k$. In Eq.~(\ref{eq:rreate}) we made the inverse function explicit since it will be used below.
Since we focus on a single rate we will drop the site dependency for all the quantities at hand.
The rates $\Gamma$ are random variables, since they depend on the difference of the random field on two contiguous sites.
The distribution of the difference $\Delta h$ can be easily extracted form the random fields' one (both $h_k$ and $h_{k+1}$ are identically distributed with the same probability between $-h$ and $h$)
\begin{equation}
p(\Delta h)=\frac{1}{4h^2}\left(2h-\left|\Delta h\right|\right),\quad \mathrm{with}\quad \Delta h\in \left[-2h,h\right].
 \label{eq:dhdistr}
\end{equation}
From this distribution the distribution of rates $P(\Gamma)$ can be defined as
\begin{equation}
P(\Gamma)=\sum_{s\in \{+,-\}}\left|\frac{d g_{V}^{-1}(\Gamma)_s}{d\Gamma}\right| p(g_{V}^{-1}(\Gamma)_s),
 \label{eq:inversion}
\end{equation}
the boundaries being 
\begin{equation}
\Gamma \in\left[\frac{1}{1+(2h+V)^2},1 \right].
 \label{eq:boundaries}
\end{equation}
The index $s$ in Eq.~(\ref{eq:inversion}) is summed over the inverse functions in the region under consideration. As depicted in Fig.~(\ref{fig:1}) the inverse function is multivalued in the region $\Delta h \in \left[-2h,2h-2V\right]$, so that in Eq.~(\ref{eq:inversion}) we have to sum on both the solutions (namely $g_{V}^{-1}(\Gamma_k)_{\pm}$). In the region $\Delta h\in \left(2h-2V,2h\right]$ on the other hand the inverse function is single-valued having as single contribution the branch $g_{V}^{-1}(\Gamma_k)_{+}$. 
The probability density for the rates Eq.~(\ref{eq:inversion}) is defined then as
\begin{equation}
P(\Gamma)=\frac{1}{8h^2\Gamma^{\frac{3}{2}}\sqrt{1-\Gamma}}
  \begin{cases}
    2h-\left|V+\sqrt{\frac{1}{\Gamma}-1}\right|  & \quad \text{if } \frac{1}{1+(V+2h)^2}\leq\Gamma< \frac{1}{1+(V-2h)^2}\\
    4h-\left|V-\sqrt{\frac{1}{\Gamma}-1}\right|-\left|V+\sqrt{\frac{1}{\Gamma}-1}\right|       & \quad \text{if } \frac{1}{1+(V-2h)^2}\leq\Gamma\leq 1.\\
    
  \end{cases}
 \label{eq:distributionrates}
\end{equation}
The joint probability $P(\Gamma_k,\Gamma')$ used in Eq.~(7) in the main text is more involved and cumbersome to write in a close form. We can though express it in a form that allows for direct numerical integration. Calling $h_k,h_{k+1},h'$ the random fields on respectively the $k$, $k+1$, and relaxed sites (see main text for an explanation) we can define

\begin{equation}
P(\Gamma_k,\Gamma')=\frac{1}{8h^3}\int_{-h}^h dh_k~dh_{k+1}~dh'\quad \delta\left(\Gamma_k-\frac{h^2}{1+(h_{k+1}-h_k)^2}\right) \quad \delta\left(\Gamma'-\frac{h^2}{1+(h'-h_{k+1})^2}\right).
 \label{eq:joint}
\end{equation}
This expression can be readily plugged in Eq.~(7) giving the results presented in Fig.~2(b) in the main text.

\end{document}